# Fuzzy Model Identification based on Mixture Distribution Analysis for Bearings Remaining Useful Life Estimation Using Small Training Data Set

Fei Huang[a, b,*], Alexandre Sava[a], Kondo H. Adjallah[a], Zhouhang Wang[a]

[a] Laboratoire de Conception, Optimisation et Modélisation des Systèmes, LCOMS EA 7306, Université de Lorraine, Metz 57000, France

[b] Huaiyin Institute of Technology, 1 Meicheng Rd, Huaian 223003, Jiangsu, P.R. China

* Corresponding author e-mail address: fei.huang@univ-lorraine.fr

**Abstract —** The research work presented in this paper proposes a data-driven modeling method for bearings remaining useful life estimation based on Takagi-Sugeno (T-S) fuzzy inference system (FIS). This method allows identifying the parameters of a classic T-S FIS, starting with a small quantity of data. In this work, we used the vibration signals data from a small number of bearings over an entire period of run-to-failure. The FIS model inputs are features extracted from the vibration signals data observed periodically on the training bearings. The number of rules and the input parameters of each rule of the FIS model are identified using the subtractive clustering method. Furthermore, we propose to use the maximum likelihood method of mixture distribution analysis to calculate the parameters of clusters on the time axis and the probability corresponding to rules on degradation stages. Based on this result, we identified the output parameters of each rule using a weighted least square estimation. We then benchmarked the proposed method with some existing methods from the literature, through numerical experiments conducted on available datasets to highlight its effectiveness.

# 1 Introduction

Condition-based maintenance (CBM) is a relevant strategy to prevent bearings failure and thus reduce downtimes in manufacturing systems [1], where rotating machinery are standard components and play an essential role. Bearing failures have significant influences on the overall performance of manufacturing systems. Hence, bearings remaining useful life (RUL) estimation could provide high-value information in CBM decision support.

Bearings degradation is a complex process as the degradation process is often different, even for identical bearings, identically loaded under similar operating conditions. Therefore, to model and identify bearings degradation process are challenging tasks. There are three main categories of approaches for estimating bearings RUL: event data based reliability approaches, condition data based prognostics approaches, and integrated approaches based on both event and condition data [2]. The event-data based reliability approaches rely on failure event records related to various degradation mechanisms, for a population of identical bearings. These approaches often use probabilistic models and need a significant amount of data for model identification [3] [4] [5]. Integrated approaches rely on both event and condition data, and they also need a significant amount of historical data for the underlying reliability models requirements. As a consequence, they are not suitable when small amounts of historical data are available.

In this paper, we address the estimation problem of bearing's RUL, from run-to-failure historical data recorded on only a small number of bearings. In practice, this situation appears with the introduction of a new type of bearings to which the proposed method may apply in a data-based condition monitoring approach. The condition data may include the bearings vibration, acoustic emission, eddy current, and voltage signals. Acoustic emission is a suitable signal for large-sized bearings under low-speed conditions [6]. Current and voltage signals are available only in the electrical devices. Vibration signals are the most widely used for bearings monitoring due to low sensor cost and ease of implementation for measurement. Therefore, in this work, we will rely on the vibration signals.

In the literature, due to the complexity of bearings degradation processes, different models have been used in various methods for estimating and forecasting bearings RUL. Thus, most of the models are composite and obtained using the bearings operating condition data. The prognostics method in [7] is data-driven and allows RUL continuous assessment through the mixture of Gaussians' hidden Markov models. The feed-forward artificial neural network (FFNN), with as input the root mean square (RMS) and the kurtosis of time fitted hazard rates, has been used in [8] to estimate and predict the RUL from bearings recent and current operating data. Also, combining relevance vector machines and exponential regression [9] allows estimating and continuous updating of bearings' RUL. In [10] the Kalman particle filter is used to model bearings stochastic degradation process, estimate and predict the RUL from their operating data. Although the results with the composite models are relatively accurate, these models require a significant amount of data for deriving and identifying the models. Our aim here is to work out a method capable of identifying a simple structure model for estimating bearings RUL upon introduction of new bearings, from a small available amount of historical data.

The fuzzy theory has demonstrated its efficiency for the cases with incomplete information [11]. Moreover, fuzzy sets based modeling techniques have the capability of modeling complex nonlinear systems by decomposing the modeling problem into some simpler linear sub-problems [12]. Among the classic fuzzy systems, the Takagi-Sugeno (T-S) Fuzzy Inference System (FIS) is computationally efficient and more suitable for mathematical analysis compared to the Mamdani FIS [13]. Thus, we propose a T-S FIS model identification method for bearings RUL estimation.

Considering that the rule-based structure of T-S FIS needs to be fixed, various methods based on automated learning have been proposed [14][15] [16]. However, with a small amount of available training data samples, the accurate model identification is not easy, for the lack of information associated with the population characteristics.

In rotating machinery, bearings, even identical, operating under cyclic and homogeneous loads, are subjected to various degradation mechanisms due to their complex structure and multi-positionality of faults. Thus, we suggest using the maximum-likelihood estimation technique of multivariate mixture distribution, which is an effective way of mixture distribution analysis [17]. A mixture distribution characterizes different operating regimes or conditions from bearings population.

Our method identifies the parameters of the T-S FIS model in two steps. It allows estimating the RUL of bearings using historical data from a complete run-to-failure cycle for a small number of identical bearings.

The rest of the paper is organized as follows. In section 2, the problem statement is presented. In section 3, we model the T-S FIS, the inputs, and the output fuzzy sets for the bearings RUL estimation. Then, Section 4 proposes an identification method of the T-S FIS through mixture distribution analysis from the input and output data. The T-S FIS, thus obtained, is implemented in Section 5, for numerical experimental studies on benchmark data. Lastly, Section 6 concludes this work and proposes some perspectives.

## 2 Problem statement

This work aims to find an effective method to identify the T-S FIS model for estimating the remaining life of bearings. The model is fed with features data extracted from vibration signals observed over a complete run-to-failure cycle on a small number of identical bearings operating under identical load. Moreover, there is no available information on the failure modes nor a fixed failure threshold. The obtained models are devised for estimating the RUL of bearings, based on a small number of bearings data. **Fig. 1** illustrates the definitions of the past useful life (PUL) and the RUL associated with the useful lifetime of bearings.

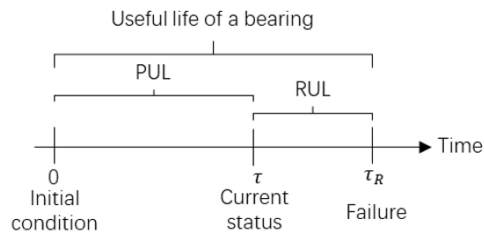

**Fig. 1.** Characteristic parameters related to the useful life of a bearing.

Let $\rho$ define the past useful life ratio of a bearing in equation (1).

$$\rho = \frac{PUL}{PUL + RUL} = \frac{\tau}{\tau_R} \quad (1)$$

where $\tau$ denotes the PUL which is known and corresponds to the period that the bearing was used until the current observation, $\tau_R$ denotes the lifetime, which is the time over a complete run-to-failure cycle of a bearing.

Using relation (1), one can calculate the RUL defined by $\Delta\tau_R = \tau_R - \tau$ based on the PUL and the past useful life ratio using (2), as it follows:

$$\Delta\tau_R = \left(\frac{1}{\rho} - 1\right)\tau \quad (2)$$

Let $V_k = [v_{k,1}, ..., v_{k,I}]^T$ denotes the input vector of the model, $k=1,2,...,K$ is the rank of observation and $i=1,2,...,I$ is the variable label. The $V_k$ vector contains the features or indicators extracted from the vibration signals data from bearings monitoring. The corresponding output of the model is the estimated past useful life ratio denoted by $\hat{\rho}_k$. **Fig. 2** depicts the general schema of the model.

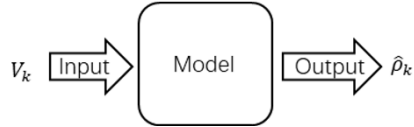

**Fig. 2.** The scheme for bearings RUL estimation.

Under this scheme, there are two phases for bearings RUL estimation: offline model identification by using training data sets, and online implementation for the bearings RUL estimation. Compared with online approaches using recurrent model identification, there are two significant merits in such kind of scheme. One is that there is no requirement for storing historical information. Another is that there is no requirement to predefine the failure threshold of bearings. For the first merit, only the current information of bearings is required under such a scheme. For the second merit, as we know, the degradation processes of bearings are diverse even for identical bearings under identical working load. The predefinition of bearings failure threshold is not easy even for identical bearings under identical working load, which seriously affects the performance of approaches for bearings RUL estimation.

Under the scheme in this work, the model yields the past useful life ratio defined in equation (1), which value ranges from 0 to 1. When the value of $\rho$ is close to 0, the bearing is at the very beginning of its whole lifetime. At this very beginning stage of its lifetime, the bearing is entirely new and healthy with top reliability. Hence, one can say the bearing at that instant is at a stage of maximum health. When the value of $\rho$ tends to 1, the bearing is at the very end of its whole lifetime. At the very end stage of its lifetime, the bearing is completely damaged. Hence, one can say the bearing at that instant is running to failure. During the training phase, the past useful life ratio $\rho$ is obtained by using equation (1), while the model allows estimating the past useful life ratio $\hat{\rho}$ for the next phases, and the equation (2) yields the RUL.

The main contributions of this work are:

1) a TS-FIS model identification approach for stochastic processes, by training from a small-size dataset, with application to bearings degradations processes.

2) a modeling and identification method of fuzzy membership functions on bearings degradation processes using the maximum likelihood estimation technique of multivariate mixture distributions.

## 3 Modeling of the T-S FIS

### 3.1 Model of input and output variables

We use the training sample data to establish the FIS model while identifying the appropriate number of rules. For that purpose:

1) Let us consider the degradation as a fuzzy set $S$ defined on a universe of discourse of degradation, and let us consider a finite number of fuzzy subsets $s_l$ of degradation levels ($S = \bigcup_l s_l$), $l \in \mathbb{N}$. A membership function $\mu_{s_l}(V)$, $l = 1, \cdots, L$, characterizes each subset, where $V$ is the vector of the feature extracted from the vibration signal and $L$ the number of degradation stages.

2) Let's consider the RUL as a fuzzy set $Z$ defined on the time universe of run-to-failure of the bearings, and let's consider a finite number of fuzzy subsets $z_l$ of $(Z = \bigcup_l z_l), l \in \mathbb{N}$. A membership function $\mu_{z_l}(\rho), l = 1, \cdots, L$, characterizes each subset, where $\rho$ is the past useful life feature corresponding to the degradation measured (**Fig. 3**).

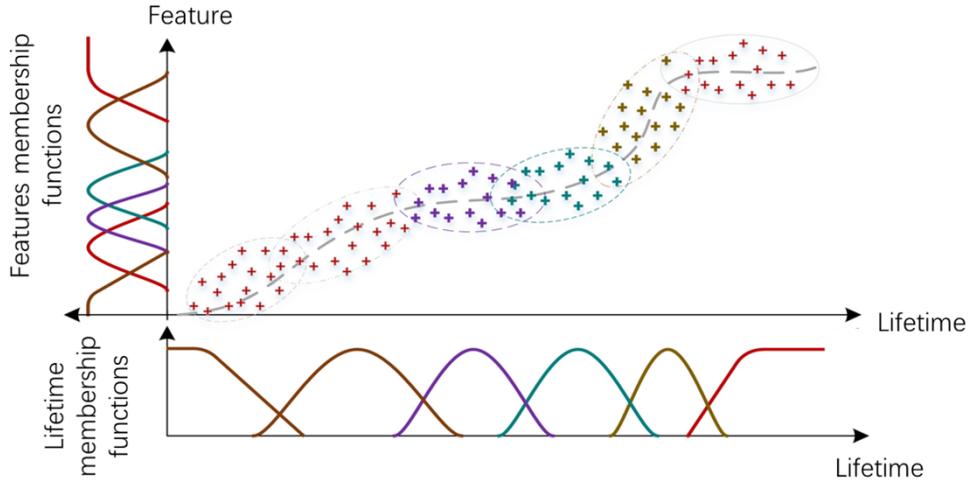

**Fig. 3.** Graph of relationships between degradation features and the lifetime feature.

**Definition**

A clustering rule is an application that associates a degradation feature input data to one, and only one fuzzy subset of degradation and the corresponding past useful life.

**Hypothesis: Degradation stage**

We assume that degradation features can reveal the bearings degradation state, to some extent, lifetime features as well as bearings RUL.

**Proposition 1** Based on the previous hypothesis, for each degradation stage, there exists a rule that associates several degradation features data to degradation state.

**Proposition 2** Based on the previous hypothesis and proposition 1, a rule can associate a datum of a fuzzy subset to only one degradation state.

**Property 1** Assume two rules $R_c$ and $R_j$ ($c \neq j$). If for the whole fuzzy set of premises, the application of the two rules yields the same results, then $R_c$ is identical to $R_j$.

**Property 2** The number of rules is equal to the number of degradation states.

**Proof:** Let us denote $L$ the number of degradation states and $J$ the number of rules.
   a) Assume $L < J$, then there exists at least one rule that associates the same data to at least two states, and so the proposition 2 is not fulfilled.
   b) Assume $L > J$, and that and the proposition 1 and are fulfilled, then there exist at least two rules $R_c$ and $R_j$ ($c \neq j$) that associate the same dataset to the same states. Then $R_c$ is identical to $R_j$.
   So, $L = J$.

### 3.2 T-S FIS modeling for bearings RUL estimation

The model of bearing degradation process, depicted in **Fig. 3**, is nonlinear. One consistent approach to identify the parameters of such a nonlinear process model consists of identifying the operating regimes first [18]. Thus, one can first establish local linear models by gathering the operating regime data, then merging the obtained continuous piece-wise local linear models to construct a nonlinear model as a succession of continuous piecewise linear models:

$$\hat{\rho}_k = \sum_{j=1}^{J} \bar{w}_j(V_k)\,(a_j V_k + b_j) \tag{3}$$

where $\bar{w}_j(V_k)$ is the activation function of the $j^{th}$ ($j=1,2…J$) local linear continuous piecewise function of the corresponding operating regime of the nonlinear model, and $[a_j, b_j]$ are the parameters of the linear function.

We chose using fuzzy sets to identify the operating regimes because of the difficulties of precise recognition of the different regimes. Thus, the nonlinear model will be represented by a T-S fuzzy inference system [19], where the rules are defined as it follows:

$R_j$: If $V_k$ is $A_j$ then the degradation is described by $(y_j, r_j)$, $j = 1, 2, \cdots, J$

where $y_j = a_j V_k + b_j$ defines the $j^{th}$ degradation continuous piece-wise local linear model, $a_j = [a_{j,1}, a_{j,2}, …, a_{j,I}]$ and $b_j$ are the parameters of the $j^{th}$ local linear model, $A_j$ is the related fuzzy subset, and $r_j \in [0, 1]$ is the weight of the rule $R_j$. The designer of the FIS usually chooses the value of $r_j$. When such knowledge is not available, let $r_j = 1$, thus, there is no effect of $r_j$ on the inference process of rules.

For a multi-input model, the antecedent part "$V_k$ is $A_j$" can be presented as a logical combination of propositions with univariate fuzzy sub-subsets defined for individual entry $v_{k,i}$ of the vector $V_k$, usually in the following conjunctive form:

$R_j$: If $v_{k,1}$ is $A_{j,1}$ and $v_{k,2}$ is $A_{j,2}$ and $v_{k,3}$ is $A_{j,3}$ and… $v_{k,I}$ is $A_{j,I}$, then $(y_j = a_j V_k + b_j, r_j)$ $j=1, 2, …, J$.

The product of membership degrees defines the degree of fulfillment of the rule by individual input membership functions $\mu_{j,i}(v_{k,i})$ and the rule's weight $r_j$ as in (4):

$$w_j(V_k) = r_j \prod_{i=1}^{I} \mu_{j,i}(v_{k,i}) \tag{4}$$

Rules aggregation allows obtaining the output (5):

$$\hat{\rho}_k = \frac{\sum_{j=1}^{J} w_j(V_k)\,(a_j V_k + b_j)}{\sum_{j=1}^{J} w_j(V_k)} \tag{5}$$

$$\bar{w}_j(V_k) = \frac{w_j(V_k)}{\sum_{j=1}^{J} w_j(V_k)} \tag{6}$$

Substituting the normalized degree of rule's fulfillment $\bar{w}_j(V_k)$ (6) into (5), yields (3).

In the next section, we introduce a method to identify the standard structure of the T-S FIS for bearings RUL estimation under the assumptions previously stated.

## 4 T-S FIS identification

We used training datasets obtained from a sample of bearings for identifying the T-S FIS model parameters. The identification task consists of determining the number of rules and the input & output variables of each rule. Thus, the T-S FIS identification process consists of two steps:

1) First, fuzzy clusters are identified in the sample space using the technique of subtractive clustering [14]. The resulting clusters are fuzzy subsets characterizing particular regimes of the input space used for feeding the T-S FIS-model. One then obtains the number $J$ of rules and the parameters of input membership functions $\mu_{j,i}(v_{k,i})$ for each rule.
2) Second, based on the mixture distribution analysis, we estimated extracted time clusters and the priori probabilities of the fuzzy subsets. Indeed, the extracted time clusters are associated with information characterizing particular degradation stages for a population of bearings. Furthermore, the estimation method of weighted least squares enables to identify the parameters of the output membership function of each fuzzy subset, using the extracted time clusters and the priori probabilities of the fuzzy subsets. The parameters thus obtained embed more information associated with the studied bearings population and lifetime of bearings.

## 4.1 Identification of the FIS-model input parameters

The clusters obtained highlight different regimes in the samples' space. For each regime, a rule is defined. In the T-S FIS, a rule defines the relationship between the fuzzy model output and the input, relating to a local model of the system for each regime [12]. So the form and overlapping of the input membership functions provide relevant information on the nonlinear character of the system [12].

Robustness to data outliers and low computational complexity characterize the subtractive clustering method used to identify the FIS model, compared to existing complex FIS model identification methods with similar accuracy [14]. Hence, we use in this work the subtractive clustering technique to identify the number $J$ of rules with the input membership functions $\mu_{j,i}(v_{k,i})$. To build the training datasets for identifying the T-S FIS model, we used the input feature vectors $V_k$ extracted from vibration signals data collected through periodic observations on a small number of bearings over an entire run-to-failure period.

The past useful life ratio corresponding to observation $k$ in the training datasets is noted $\rho_k$.

We also used $K$ successive observation samples to train the model. These observations are gathered in a $K \times (I + 1)$ matrix $G$ (7). Each row of the $G$ matrix corresponds to an observation in the training datasets. The first $I$ columns correspond to the input, and the last column is the output of the training sample. Hence, the $G$ matrix gathers the input and output data from the training sample.

$$G = \begin{bmatrix} v_{1,1} & v_{1,2} & \cdots & v_{1,I} & \rho_1 \\ v_{2,1} & v_{2,2} & \cdots & v_{2,I} & \rho_2 \\ \vdots & \vdots & \ddots & \vdots & \vdots \\ v_{K,1} & v_{K,2} & \cdots & v_{K,I} & \rho_K \end{bmatrix} \qquad (7)$$

Each rule corresponds to a specific approximated linear regime and can be associated with a cluster in the sample space. Then, we used the subtractive clustering method [14] to identify the clusters based on the training samples in the rows of the matrix $G$. A rule relates each regime to a specific obtained cluster, and the number $J$ of obtained clusters is the number of the rules according to Property 2.

The subtractive clustering method is applied to the $G$ matrix, the collection of the input and output data from the training sample. The obtained clusters essentially reveal the prototypical structures of the input and output data space, namely the characteristic behaviors of the bearings degradation. In other words, they correspond to the degradation states of bearings. All the rules are established on the obtained clusters. Naturally, the rules are to relate to the degradation states of bearings.

The matrix $D$, given in equation (8), memorizes the centroids of the clusters. The centroid of the $j^{th}$ cluster in the input space constitutes the first $I$ elements of the $j^{th}$ row, while $c_j^*$ is the centroid of the $j^{th}$ cluster in the output space. The Gaussian function defined in equation (9) is highly employed in the literature to model the input membership functions in various application fields due to its invariance property under multiplication [20]. Therefore, we chose Gaussian functions to model the inputs membership functions.

To let (9) be consistent with the potentiality measurement of data point for being a cluster in the subtractive clustering method [14], the parameter $\sigma_i$ of the input membership function of the $i^{th}$ input feature is defined in equation (10).

$$D = \begin{bmatrix} c_{1,1} & c_{1,2,} & \cdots & c_{1,I} & c_1^* \\ c_{2,1} & c_{2,2} & \cdots & c_{2,I} & c_2^* \\ \vdots & \vdots & \ddots & \vdots & \vdots \\ c_{J,1} & c_{J,2} & \cdots & c_{J,I} & c_J^* \end{bmatrix} \qquad (8)$$

$$\mu_{j,i}(v_{k,i}) = e^{\frac{-(v_{k,i}-c_{j,i})^2}{2\sigma_i^2}} \qquad (9)$$

$$\sigma_i = r_a(max(V_i) - min(V_i))/2\sqrt{2} \qquad (10)$$

where $j=1,2,...,J$ is the label of the rules, $i=1,2...,I$ is the feature label of the input, $V_i$ is the $i^{th}$ input feature, $r_a$ is the influence constant of subtractive clustering.

We follow the suggestions in [14] to fix the acceptance threshold and rejection threshold of subtractive clustering. Additionally, we set the influence constant-coefficient $r_a$ as 0.5 and subtract constant-coefficient $r_b$ as $1.25r_a$ in this work, which results in better performance of the model compared with the settings suggested in [14].

## 4.2 Identification of the FIS-model output parameters

The output membership function parameters in each rule are identified using a training sample dataset. The sample datasets derive from the bearing degradation monitoring.

Due to the complexity of the bearing degradation process, we assume the sample datasets $V_k$ are derived from a finite and countable mixture multivariate distribution $f(V_k)$, which is a mixture of multivariate normal distributions $g_j(V_k; C_j, H_j)$ by taking proportions $\varsigma_j$. Each multivariate normal distribution is associated with a specific regime component in the sample space. We define the mixture distribution density function $f(V_k)$ by equations (11) (12) (13).

$$f(V_k) = \sum_{j=1}^{J} \varsigma_j \, g_j(V_k; C_j, H_j) \tag{11}$$

$$\sum_{j=1}^{J} \varsigma_j = 1 \tag{12}$$

$$g_j(V_k; C_j, H_j) = \frac{1}{(2\pi)^{I/2} |H_j|^{1/2}} e^{(-\frac{1}{2}(V_k - C_j) H_j^{-1} (V_k - C_j)^T)} \tag{13}$$

$$P(\theta_{R_j} | V_k) = \frac{P(\theta_{R_j}) P(V_k | \theta_{R_j})}{P(V_k)} = \frac{\varsigma_j g_j(V_k | C_j, H_j)}{f(V_k)} \tag{14}$$

where $\theta_{R_j}$ denotes the $j^{th}$ regime component associated with the $j^{th}$ component distribution, $C_j$ and $H_j$ are the covariance matrix and the mean vector associated with the $\theta_{R_j}$ regime component. The probability of membership $P(\theta_{R_j} | V_k)$ is defined in (14) [17], according to the Bayesian rule.

The $f(V_k)$ can be considered as a function with $\varsigma_j$, $H_j$ and $C_j$ as its parameters.

The maximum-likelihood estimation of the multivariate mixture distribution is an effective method for mixture distribution analysis [17]. Based on (11)(12)(13) and using the Lagrange multiplier method, the problem of multivariate mixture distribution parameters estimation with the maximum-likelihood is formulated as

$$\begin{aligned}\hat{\varsigma}_j &= \underset{\varsigma_j}{argmax}(\sum_{k=1}^{K} \ln f(V_k) - \lambda(\sum_{j=1}^{J} \varsigma_j - 1)) \\ &= \underset{\varsigma_j}{argmax}(\sum_{k=1}^{K} \ln \sum_{j=1}^{J} \varsigma_j \, g_j(V_k; C_j, H_j) - \lambda(\sum_{j=1}^{J} \varsigma_j - 1))\end{aligned} \tag{15}$$

$$\begin{aligned}\hat{\sigma}^2_{j,i} &= \underset{\sigma^2_{j,i}}{argmax}(\sum_{k=1}^{K} \ln f(V_k) - \lambda(\sum_{j=1}^{J} \varsigma_j - 1)) \\ &= \underset{\sigma^2_{j,i}}{argmax}(\sum_{k=1}^{K} \ln \sum_{j=1}^{J} \varsigma_j \, g_j(V_k; C_j, H_j) - \lambda(\sum_{j=1}^{J} \varsigma_j - 1))\end{aligned} \tag{16}$$

$$\begin{aligned}\hat{c}_{j,i} &= \underset{c_{j,i}}{argmax}(\sum_{k=1}^{K} \ln f(V_k) - \lambda(\sum_{j=1}^{J} \varsigma_j - 1)) \\ &= \underset{c_{j,i}}{argmax}(\sum_{k=1}^{K} \ln \sum_{j=1}^{J} \varsigma_j \, g_j(V_k; C_j, H_j) - \lambda(\sum_{j=1}^{J} \varsigma_j - 1))\end{aligned} \tag{17}$$

where $\lambda$ is the Lagrange multiplier, $\hat{c}_{j,i}$ and $\hat{\sigma}^2_{j,i}$ are the maximum-likelihood estimated mean and variance of the $i^{th}$ feature of $V$ associated with $\theta_{R_j}$, which are respectively the element of $H_j$ and $C_j$.

Since the mixture distribution $f(V_k)$ is a mixture of multivariate normal distributions, the function in the brackets of the right side of equations (15) (16) and (17) is convex. Hence, when $\varsigma_j$ takes the maximum-likelihood value $\hat{\varsigma}_j$, the equation (18) is satisfied. Similarly, when $\sigma^2_{j,i}$ takes the maximum-likelihood value $\hat{\sigma}^2_{j,i}$, the equation (19) is satisfied, when $c_{j,i}$ takes the maximum-likelihood value $\hat{c}_{j,i}$, the equation (20) is satisfied.

$$\frac{\partial(\sum_{k=1}^{K} \ln \sum_{j=1}^{J} \varsigma_j \, g_j(V_k; C_j, H_j) - \lambda(\sum_{j=1}^{J} \varsigma_j - 1))}{\partial \varsigma_j} = 0 \tag{18}$$

$$\frac{\partial(\sum_{k=1}^{K} \ln \sum_{j=1}^{J} \varsigma_j \, g_j(V_k; C_j, H_j) - \lambda(\sum_{j=1}^{J} \varsigma_j - 1))}{\partial \sigma^2_{j,i}} = 0 \tag{19}$$

$$\frac{\partial(\sum_{k=1}^{K} \ln \sum_{j=1}^{J} \varsigma_j \, g_j(V_k; C_j, H_j) - \lambda(\sum_{j=1}^{J} \varsigma_j - 1))}{\partial c_{j,i}} = 0 \tag{20}$$

Under the condition of components distribution with unequal covariance matrices, according to the derivations respectively on (18) (19) (20) using (14) in [17], the maximum-likelihood estimated values of parameters in a mixture of multivariate normal distributions can be obtained as it follows:

$$\hat{\varsigma}_j = \frac{1}{K} \sum_{k=1}^{K} \hat{P}(\theta_{R_j}|V_k) \tag{21}$$

$$\hat{c}_{j,i} = \frac{\sum_{k=1}^{K} x_{k,i} \hat{P}(\theta_{R_j}|V_k)}{\sum_{k=1}^{K} \hat{P}(\theta_{R_j}|V_k)} \tag{22}$$

$$\hat{\sigma}^2_{j,i} = \frac{\sum_{k=1}^{K}(v_{k,i} - \hat{c}_{j,i})^2 \hat{P}(\theta_{R_j}|V_k)}{\sum_{k=1}^{K} \hat{P}(\theta_{R_j}|V_k)} \tag{23}$$

According to the definition in formula (6), the normalized rule degree of fulfillments $\bar{w}_j(V_k)$ are obtained for the following conditions:

$$\bar{w}_j(V_k) \in [0,1], \forall j, k; \sum_{j=1}^{J} \bar{w}_j(V_k) = 1, \forall k; \; 0 < \sum_{k=1}^{K} \bar{w}_j(V_k) < K, \forall j.$$

Considering the $j^{th}$ regime and the probability of membership definition in eq. (14), $\bar{w}_j(V_k)$ can be considered as the estimated value of the probability of membership $P(\theta_{R_j}|V_k)$. Hence, $\bar{w}_{j,k} = \hat{P}(\theta_{R_j}|V_k)$. Moreover, according to the formulas (14), $\hat{P}(\theta_{R_j}) = \hat{\varsigma}_j$. We can rewrite equations (21)(22)(23) as follows:

$$\hat{P}(\theta_{R_j}) = \frac{1}{K} \sum_{k=1}^{K} \hat{P}(\theta_{R_j}|V_k) = \frac{1}{K} \sum_{k=1}^{K} \bar{w}_j(V_k) \tag{24}$$

$$\hat{c}_{j,t} = \frac{\sum_{k=1}^{K} \tau_k \hat{P}(\theta_{R_j}|V_k)}{\sum_{k=1}^{K} \hat{P}(\theta_{R_j}|V_k)} = \frac{\sum_{k=1}^{K} \tau_k \bar{w}_j(V_k)}{\sum_{k=1}^{K} \bar{w}_j(V_k)} \tag{25}$$

$$\hat{\sigma}^2_{j,t} = \frac{\sum_{k=1}^{K}(\tau_k - \hat{c}_{j,t})^2 \hat{P}(\theta_{R_j}|V_k)}{\sum_{k=1}^{K} \hat{P}(\theta_{R_j}|V_k)} = \frac{\sum_{k=1}^{K}(\tau_k - \hat{c}_{j,t})^2 \bar{w}_j(V_k)}{\sum_{k=1}^{K} \bar{w}_j(V_k)} \tag{26}$$

where $\tau_k$ is the time-coordinate of the input $V_k$ associated with the corresponding bearing past useful life, $\hat{c}_{j,t}$ and $\hat{\sigma}^2_{j,t}$ are the maximum-likelihood estimated values of the time-coordinate centroid and the variance associated with $\theta_{R_j}$. According to equations (25) and (26), one can consider that the parameters $\hat{c}_{j,t}$ and $\hat{\sigma}^2_{j,t}$ describe the projection of the probability distribution of the $V_k$ input onto the time-coordinate.

Following the approach in Section 4.1, we construct the input membership function for the time-coordinates in each rule of the T-S FIS model. Due to its invariance property under multiplication [20], we use the Gaussian type membership function for the time-coordinates, the same with the input space. Hence, the input membership function characterizing the time clusters is:

$$\hat{\mu}_{j,t}(\tau_k) = e^{\frac{-(\tau_k - \hat{c}_{j,t})^2}{2\hat{\sigma}_{j,t}^2}} \tag{27}$$

Now, we rewrite the formula (4) (6) and (3),

$$\widetilde{w}_j(V_k) = \hat{P}(\theta_{R_j})\hat{\mu}_{j,t}(\tau_k) \prod_{i=1}^{I} \mu_{j,i}(v_{k,i}) = \hat{P}(\theta_{R_j})\hat{\mu}_{j,t}(\tau_k) w_j(V_k) \tag{28}$$

$$\bar{\widetilde{w}}_j(V_k) = \frac{\widetilde{w}_j(V_k)}{\sum_{j=1}^{J} \widetilde{w}_j(V_k)} \tag{29}$$

$$\hat{\rho}_k = \sum_{j=1}^{J} \bar{\widetilde{w}}_j(V_k) (a_j^T V_k + b_j) \tag{30}$$

In the left-hand side of the second equal sign of formula (28), the probability $\hat{P}(\theta_{R_j})$ associated with the population is obtained by equation (24), also called the priori probability, which can be considered as the confidence associated with the corresponding rule. In parallel, $\hat{\mu}_{j,t}(\tau_k)$ in the formula (28) can be considered as an individual membership degree of an extra dimension of input features compared with (4). According to the right-hand side of the second equal sign of formula (28), $\tilde{w}_j(V_k)$ is $w_j(V_k)$ weighted by $\hat{P}(\theta_{R_j})\hat{\mu}_{j,t}(\tau_k)$. Comparing with $\bar{w}_j(V_k)$ in (6), $\tilde{\bar{w}}_j(V_k)$ in (29) brings extra information associated with the population and lifetime of bearings. Since $\hat{P}(\theta_{R_j})$ and $\hat{\mu}_{j,t}(\tau_k)$, especially $\hat{\mu}_{j,t}(\tau_k)$, are derived from the estimation process, to obtain $\tilde{\bar{w}}_j(V_k)$, there is no requirement to add more features to the input of the T-S FIS model. In other words, there is no requirement to increase the number of input membership functions in the antecedent parts of rules.

For the sake of simplicity, let us define the following matrix and vectors:

$$\Phi = \begin{bmatrix} \tilde{\bar{w}}_1(V_1)V_1^T & \tilde{\bar{w}}_1(V_1) & \tilde{\bar{w}}_2(V_1)V_1^T & \tilde{\bar{w}}_2(V_1) & \cdots & \tilde{\bar{w}}_J(V_1)V_1^T & \tilde{\bar{w}}_J(V_1) \\ \tilde{\bar{w}}_1(V_2)V_2^T & \tilde{\bar{w}}_1(V_2) & \tilde{\bar{w}}_2(V_2)V_2^T & \tilde{\bar{w}}_2(V_2) & \cdots & \tilde{\bar{w}}_J(V_2)V_2^T & \tilde{\bar{w}}_J(V_2) \\ \vdots & \vdots & \vdots & \vdots & \ddots & \vdots & \vdots \\ \tilde{\bar{w}}_1(V_K)V_K^T & \tilde{\bar{w}}_1(V_K) & \tilde{\bar{w}}_2(V_K)V_K^T & \tilde{\bar{w}}_2(V_K) & \cdots & \tilde{\bar{w}}_J(V_K)V_K^T & \tilde{\bar{w}}_J(V_K) \end{bmatrix}$$

$$\vec{\rho} = [\rho_1 \quad \rho_2 \quad \cdots \quad \rho_K]^T$$
$$\hat{\vec{\rho}} = [\hat{\rho}_1 \quad \hat{\rho}_2 \quad \cdots \quad \hat{\rho}_K]^T$$
$$\beta = [a_1 \quad b_1 \quad a_2 \quad b_2 \quad \cdots \quad a_J \quad b_J]^T.$$

Hence, for the entire training sample, formula (30) can be rewritten as follows:

$$\hat{\vec{\rho}} = \Phi\beta \tag{31}$$

For the training data, the left side and the first matrix of the right side are constants. The parameters are in the second matrix of the right side. Thus, to determine the parameters of all the local models, it suffices to find the vector $\beta$, which minimizes $\|\vec{\rho} - \Phi\beta\|^2$, i.e., to solve the following problem using the linear least-squares estimation technique:

$$\hat{\beta} = \underset{\beta}{argmin}\|\vec{\rho} - \Phi\beta\|^2 \tag{32}$$

Since we use the maximum-likelihood estimated values to obtain the $\tilde{\bar{w}}_j(V_k)$ which is weighted, the linear least-squares estimation is weighted, and the solutions embed more information associated with the population and lifetime of bearings, compared with using $\bar{w}_j(V_k)$. On the other hand, the method avoids increasing the number of input membership functions to complicate the structure of the T-S FIS model.

We use the Moore-Penrose inverse [21] for minimizing (32) to obtain $\hat{\beta}$ as in (33).

$$\hat{\beta} = (\Phi^T\Phi)^{-1}\Phi^T\vec{\rho} \tag{33}$$

### 4.3 Algorithm for the T-S FIS model identification

The identification process proposed in this paper summarizes as hereafter (**Fig. 3**).

---

Identifying the input parameters
1. Set the Gaussian input membership function of T-S FIS in equation (9).
2. Compute the matrix $D$ in equation (8).
3. Determine the rules number $J$ and the parameter $c_{j,i}$ according to matrix $D$.
4. Compute the parameters $\sigma_i$ using equation (10).

---

Identifying the output parameters
1. According to (24), estimate the probability of a given degradation stage $\hat{P}(\theta_{R_j})$.
2. Project the feature space clusters on the time-space, with $\hat{\mu}_{j,t}$ obtained using (27).
3. Estimate $\tilde{\bar{w}}_j(V_k)$ using (29).
4. Calculate the output parameters $a_j$ and $b_j$ using (33).

---

The parameters $J$, $c_{j,i}$, $\sigma_i$, $a_j$ and $b_j$ obtained during the identification process characterize the T-S FIS model for the bearings RUL estimation. Additionally, the parameters $r_j$ are set to 1.

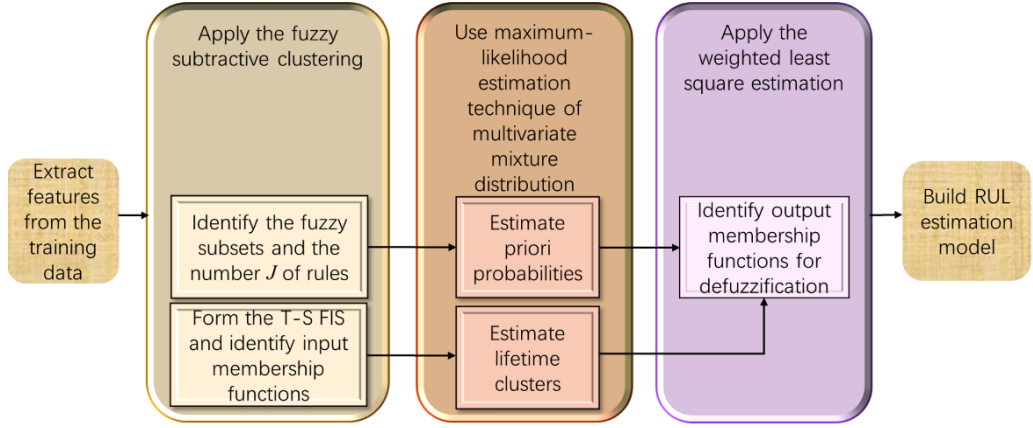

**Fig. 4.** The flow diagram of the proposed algorithm for the T-S FIS model identification.

## 5 Case study

The effectiveness of the T-S FIS identification method introduced in this paper for bearings RUL estimation is assessed, using respectively two benchmark datasets. One benchmark is related to ball bearings, while the other is related to double row bearings. We have selected the following features for the input of our model in this paper. The root mean square (RMS), derived from vibration signals data, provides additional information on the energy quantity of the signal. Let us recall that RMS is a critical time-domain feature used to estimate bearing degradation [22]. The spectral entropy (SE) is a frequency domain feature for bearings degradation assessment [23]. SE can provide a measure of vibration signals' spectral power distribution. The approximate entropy (AE) [24], the largest Lyapunov exponent (LLE) [25], and the correlation dimension (CD) [26], developed based on the chaos theory, are three measurement features of the phase-space dissimilarity for bearings condition monitoring and prognosis [27]. Each value of each feature is obtained from the vibration signal in one sample interval. By processing the vibration signal in $k^{th}$ sample interval, we shall denote $RMS_k$ the RMS value, $SE_k$ the SE value, $AE_k$ the AE value, $LLE_k$ the LLE value, $CD_k$ the CD value. According to the degradation index method in [28], we shall denote $DIAE_k$ the degradation index of $AE_k$. In the current case study, the model inputs to identify the T-S FIS are chosen among these six above features.

We use equations (1) and (2) to calculate the RUL based on the estimated past useful life ratio which is the model output.

In general, there are three main categories of identification methods for T-S FIS addressed in this work, e.g. methods based respectively on clustering algorithms, neural networks and metaheuristic [29]. The identification method introduced in [14] is based on the subtractive clustering and least square estimation (FSC-LSE), which is a representative of the clustering based identification methods category. The method introduced in [20] is based on the least square estimation and backpropagation gradient descent (BPGD-LSE), which is a representative of the neural networks based identification methods category. Among the metaheuristic algorithms, the genetic algorithm (GA) [30] is the famous population generation based bioinspired metaheuristic algorithm, and the simulated annealing (SA) [31] is one typical single solution based metaheuristic algorithm. For convenience, in the sequel, we shall denote A the FSC-LSE based method, B the BPGD-LSE based method, C the GA based method, and D the SA based method. Also, we shall denote E the method proposed in this paper, based on subtractive clustering with maximum likelihood and weighted least square estimation (MLWLSE).

In this section, the proposed method E is benchmarked with four existing methods for T-S FIS identification, e.g. method A, B, C, and D.

The setting for the subtractive clustering in method A is performed as described earlier. For method B, the training epoch number is set to 30 steps. The initial step size, step size decrease rate, and step size increase rate for method B are set as suggested in

[20]. For the setting of the GA, the population size is set to 200, the mutation function is adaptive feasible function, crossover fraction is 0.8, and the maximum number of generations is 30. For the SA main parameters setting, the initial temperature is set to 100, the maximum number of iterations is set to 30, the reannealing interval is set to 100.

Also, let us notice that:
1) The identification results of method A is used as initial FIS for the method B, C, and D. The least-square estimation is a non-parametric method, and the backpropagation gradient descent method in method B does not affect the rules. In methods C and D presented, the metaheuristic algorithms only optimize the FIS with parameters of the input and output membership functions but not with the rules. Hence, all models identified by different methods have a consistent structure. Moreover, for methods B, C, and D, the maximum numbers of epochs/ iterations/generations are all set to 30, so to benchmark the results quickly.
2) The models identified by three methods have the same type of input memberships functions (gaussian type), the same number of rules, and the same value of $r_j = 1$.
3) All operations are implemented using the MATLAB 2016a software, on a Dell OptiPlex 7040 computer equipped with Intel Core i5-6600 CPU and 8GB RAM, operating under Windows 10 professional OS environment.

We use the relative root mean square error (RRMSE) (34) and the average of RRMSE (ARRMSE) (35) to evaluate the models' performance.

$$RRMSE_q = \sqrt{\frac{1}{K^q} \sum_{k^q=1}^{K^q} \left(\frac{\rho_{k^q} - \hat{\rho}_{k^q}}{\rho_{k^q}}\right)^2} \quad (34)$$

$$ARRMSE = \frac{1}{Q} \sum_{q=1}^{Q} RRMSE_q \quad (35)$$

where $q=1,2,...Q$ is the test bearings label, $k^q=1,2,...K^q$ is the rank of $q^{th}$ test bearing observation.

## 5.1 Case study with IEEE PHM 2012 datasets (bearing type: ball bearing)

The IEEE PHM 2012 datasets from the PRONOSTIA testbed [32] are obtained from the same bearings type subjected to three different load conditions. There is no information about failure modes and fixed failure threshold. For our study, according to [32], two ball bearings with the tags (1-1, 1-2) were used for training bearings submitted to the load condition 1, and five ball bearings (1-3, 1-4, 1-5, 1-6, 1-7) were used as test bearings submitted to the same load condition.

The bearings under condition 1 operated at 1800 rpm with 4000 N radial load. Vibration signal data were collected on each of the bearings over an entire run-to-failure period. Two accelerometers have allowed measuring the vibrations in the vertical and horizontal directions.

Data were sampled at a 25.6 kHz sampling rate and 0.1s recording duration with 10s sample intervals. Thus, each sample interval is made of 2560 measures. **Fig. 5** displays the sample parameters.

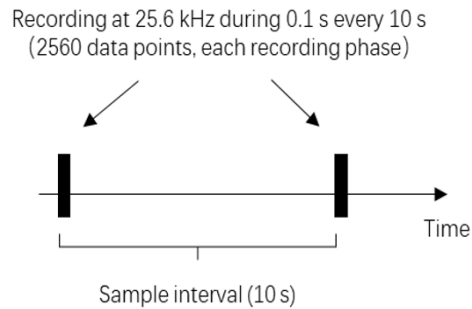

**Fig. 5.** Parameters of the sampling process in IEEE PHM 2012 datasets [32].

For simplicity, we consider only the vibration signals in the horizontal direction.

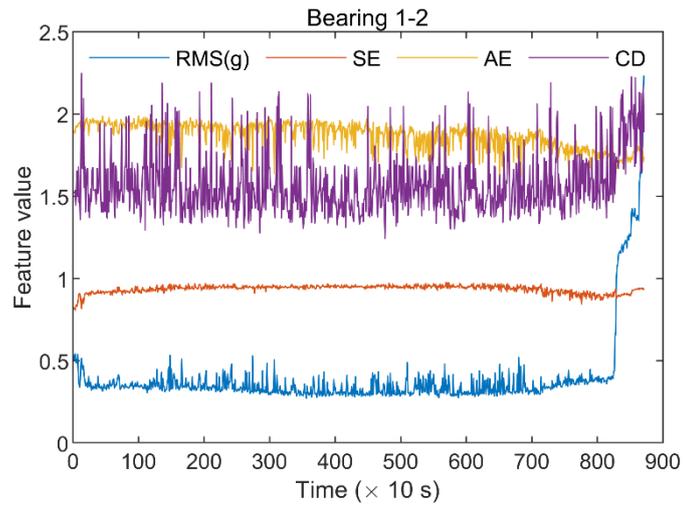

**Fig. 6** and **Fig. 7** show the RMS, SE,AE, LLE, and CD evolution graph over the whole lifecycle of the bearing 1-2.

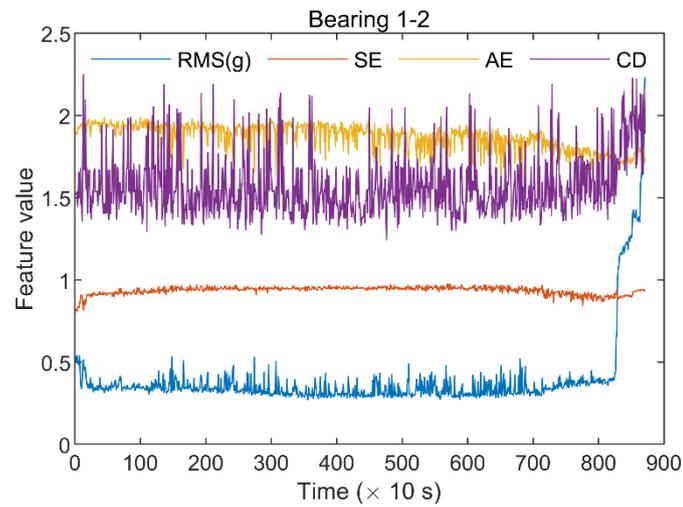

**Fig. 6.** RMS, SE,AE, and CD features over an entire run to failure period of bearing1-2.

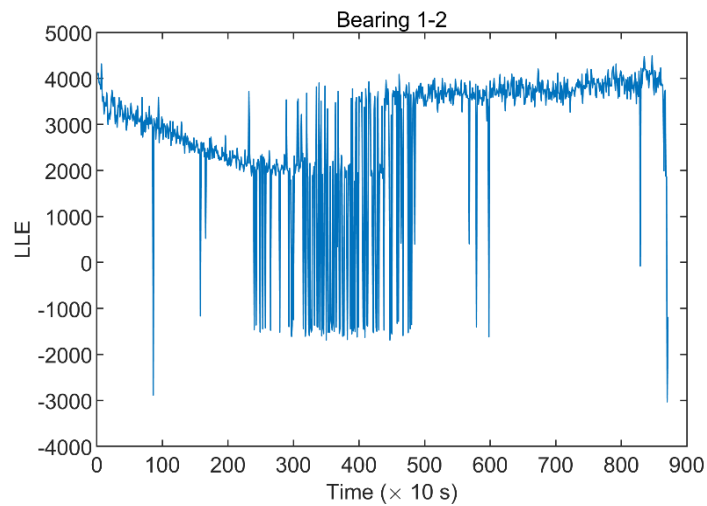

**Fig. 7.** The LLE feature over an entire run to failure period of bearing1-2.

As mentioned previously, the T-S FIS model is established based on two training ball bearings 1-1 and 1-2 and then assessed using the test ball bearings 1-3, 1-4, 1-5, 1-6, 1-7.

**Table 1**

By using IEEE PHM 2012 datasets, the results based on ($V_k=RMS_k$) RMS as input.

| Identification method | RRMSE | | | | | ARRMSE | Execution time |
|---|---|---|---|---|---|---|---|
| | Test bearing | | | | | | |
| | 1-3 | 1-4 | 1-5 | 1-6 | 1-7 | | |
| Method A (FSC-LSE) | 1.2362 | 1.5269 | 1.2433 | 1.4719 | 1.3047 | 1.3566 | 2.4219s |
| Method B (BPGD-LSE) | 1.2216 | 1.4544 | 1.2391 | 1.4600 | 1.2987 | 1.3348 | 2.7031s |
| Method C (GA) | 1.2928 | 1.4564 | 1.3046 | 1.4559 | 1.3476 | 1.3715 | 239.5469s |
| Method D (SA) | 1.6325 | 1.7260 | 1.6302 | 1.7218 | 1.6591 | 1.6739 | 5.6094s |
| Method E (MLWLSE) | **0.6979** | **0.8263** | **0.8106** | **0.8556** | **0.7991** | **0.7979** | 2.4844s |

**Table 2**

By using IEEE PHM 2012 datasets, the results based on ($V_k=[RMS_k, SE_k, AE_k, LLE_k, CD_k]^T$) 5 kinds of features as input.

| Identification method | RRMSE | | | | | ARRMSE | Execution time |
|---|---|---|---|---|---|---|---|
| | Test bearing | | | | | | |
| | 1-3 | 1-4 | 1-5 | 1-6 | 1-7 | | |
| Method A (FSC-LSE) | 0.8939 | 0.5819 | 0.7837 | 1.1346 | 0.7752 | 0.8339 | 3.1719s |
| Method B (BPGD-LSE) | 0.8657 | 0.9527 | 0.7137 | 0.8153 | 0.5998 | 0.7894 | 5.7344s |
| Method C (GA) | 0.8949 | 0.5761 | 0.8876 | 1.1057 | 1.0797 | 0.9088 | 1043.9656s |
| Method D (SA) | 1.0634 | 0.7764 | 0.9535 | 1.2657 | 1.2670 | 1.0652 | 12.5938s |
| Method E (MLWLSE) | **0.6436** | **0.5697** | **0.6879** | **1.0032** | **0.5283** | **0.6865** | 3.3906s |

According to equation (2), the estimated RUL is obtained based on the output $\hat{\rho}$ of the models. The RUL estimations during the whole lifecycle of bearing 1-3 are plotted in **Fig. 8**. The Savitzky-Golay filter of order two is applied to smooth the RUL estimations to allow decision-making without false alarms. Savitzky-Golay filter is a digital filter suitable for smoothing the equally spaced data, based on convolution process by running least-squares polynomial fitting [33]. The frame length of it is set to 61. The smoothed RUL estimations during the whole lifecycle of bearing 1-3 are shown in **Fig. 9**.

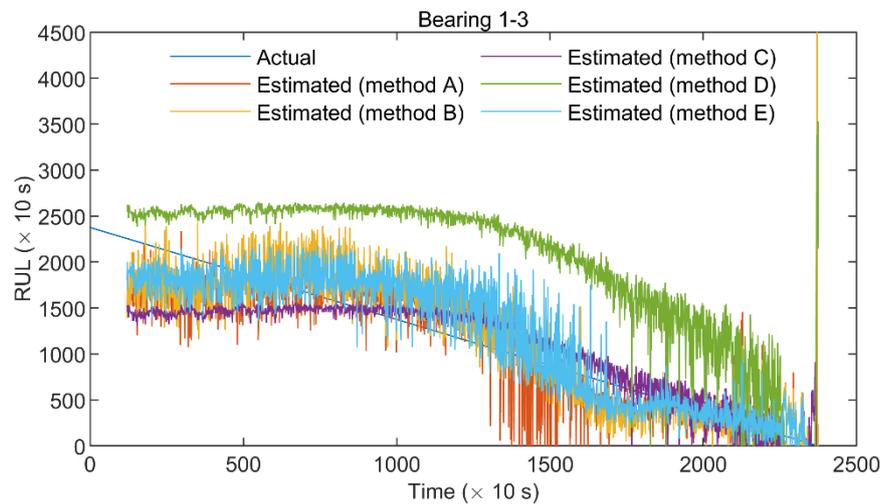

**Fig. 8.** The RUL estimations based on the input ($V_k=[RMS_k, SE_k, AE_k, LLE_k, CD_k]^T$).

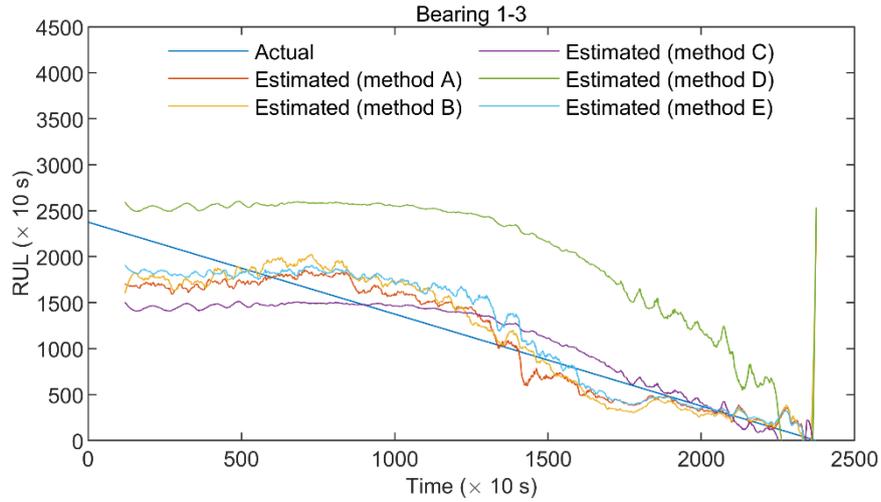

**Fig. 9.** The smoothed RUL estimations of bearing 1-3 by using Savitzky-Golay filter.

We considered two cases for IEEE PHM 2012 ball bearing dataset. First, only RMS is used as input to the model. Then, we considered five features, i.e. RMS, SE, AE, LLE, and CD, for the model input. The numerical results reveal that using a set of features provides better performance than using only one feature. The accuracy of the models identified by our method overperform, in the benchmarking, that of the other four methods for this dataset.

**5.2  Case study with IMS bearing datasets (bearing type: double row bearing)**

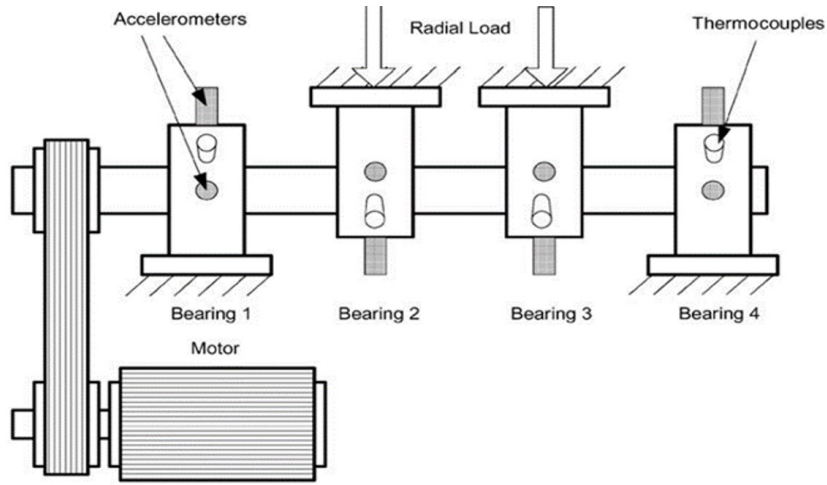

**Fig. 10.** Bearing test rig used for data acquisition of IMS bearing datasets [34].

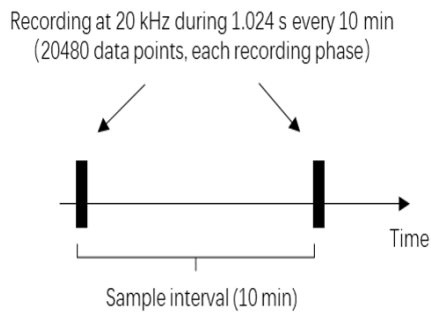

**Fig. 11.** Illustration of sampling process of data in IMS bearing datasets.

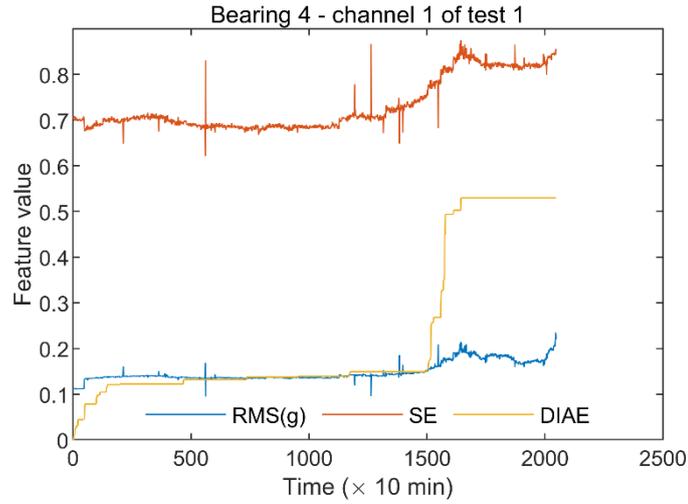

**Fig. 12.** The features over an entire run to failure period of bearing 4-channel 1 of test 1.

The IMS bearing datasets were collected by the NSF I/UCR Center for Intelligent Maintenance Systems (IMS – www.imscenter.net) with support from Rexnord Corp. in Milwaukee, WI [34]. In this dataset, three bearing run-to failure tests were performed at 2000 rpm under 6000 lbs. (26688 N) radial load on a specially designed test rig which is shown in **Fig. 10**. For each test, four Rexnord ZA-2115 double row bearings were installed on the shaft of the test rig (**Fig. 11**). The data sampling rate is 20 kHz and the data length is 20480 points in each sample interval. Among the bearings which occurred defeat at the end of tests, we choose first channel data from double row bearing 4 of the first test (**Fig. 12**) as the training set and the data from double row bearing 1 of the second test as the test set. To fit within the assumption of this work, we ignore the information about failure modes provided in the datasets. Moreover, there is no information about fixed failure thresholds.

**Table 3**

By using IMS bearing datasets, the results based on ($V_k=RMS_k$) RMS as input.

| Identification method | RRMSE | Execution time |
|---|---|---|
|  | Test bearing |  |
| Method A (FSC-LSE) | 2.1027 | 1.1875s |
| Method B (BPGD-LSE) | 2.0255 | 1.3438s |
| Method C (GA) | 1.3881 | 103.7188s |
| Method D (SA) | 1.5071 | 3.1503s |
| Method E (MLWLSE) | 0.9683 | 1.2188s |

**Table 4**

By using IMS bearing datasets, the results based on ($V_k=[RMS_k, SE_k, DIAE_k]^T$) 3 kinds of features as input.

| Identification method | RRMSE | Execution time |
|---|---|---|
|  | Test bearing |  |
| Method A (FSC-LSE) | 8.1162 | 1.3563s |
| Method B (BPGD-LSE) | 1.7420 | 2.2625s |
| Method C (GA) | 0.9229 | 269.0001s |
| Method D (SA) | 1.2669 | 4.6094s |
| Method E (MLWLSE) | 0.5892 | 1.5156s |

According to equation (2), the estimated RUL is obtained based on the output $\hat{\rho}$ of the models. The RUL estimations during the whole lifecycle of bearing 1 of test 2 are plotted in **Fig. 13**. The Savitzky-Golay filter of order two is applied to smooth the RUL estimations to allow decision-making without false alarms. Savitzky-Golay filter is a digital filter suitable for smoothing the equally spaced data, based on convolution process by running least-squares polynomial fitting [33]. The frame length of it is set to 61. The smoothed RUL estimations during the whole lifecycle of bearing 1 of test 2 are shown in **Fig. 14**.

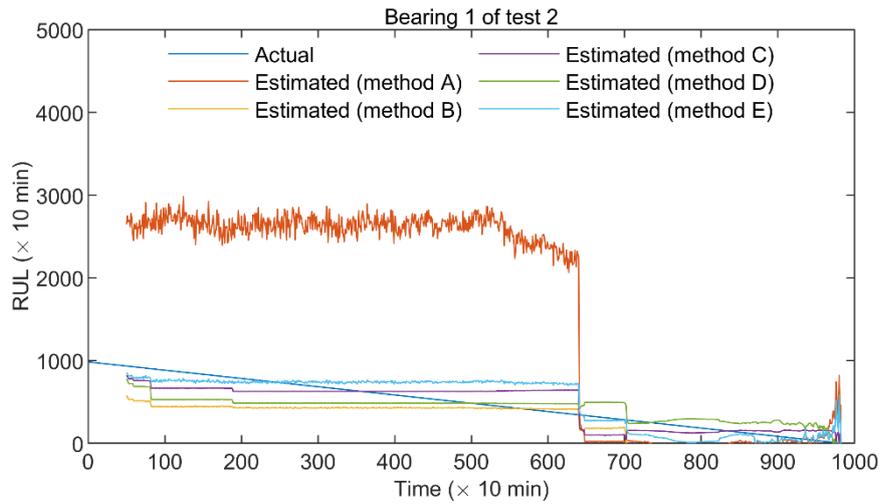

**Fig. 13.** The RUL estimations of bearing 1 of test 2 based on the input ($V_k$=[$RMS_k$, $SE_k$, $DIAE_k$]$^T$).

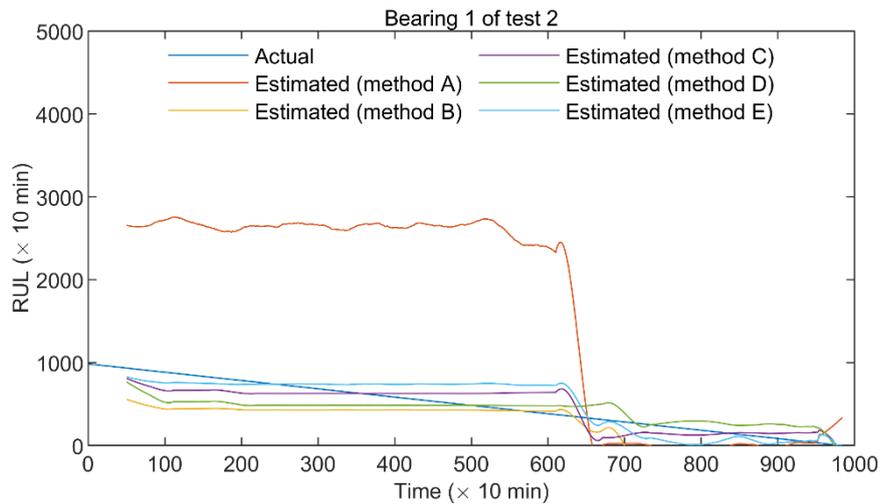

**Fig. 14.** The smoothed RUL estimations of bearing 1 of test 2 by using Savitzky-Golay filter.

We considered two cases for IMS double row bearing dataset. First, only RMS is used as input to the model. Then, we considered three features, i.e. RMS, SE, and DIAE, for the model input. The conclusions drawn for the IEEE PHM 2012 benchmark dataset are confirmed by the results obtained for the IMS bearings dataset. The numerical results show that the method proposed in this work provides a promising identification method of the T-S FIS model for bearing RUL estimation using small size datasets.

### 5.3 Summary of numerical results

Based on the same input, according to the results in Table 1 to 4:
1) Except method A, based on all the methods, the results are always improved by considering multi-input features for the degratdation model. This is because each additional feature brings complementary information for the degradation model identification.
2) Method C and D consume more execution time to yields acceptable results.
3) Method A consumed the least time to identify T-S FIS model among the five methods.

4) Method B and E consume extra execution time while obtaining models that are more accurate compared with method A. Moreover, method E is computationally effective with more accurate results, relatively to Method B.
5) Method E is the one that identifies the most accurate T-S FIS model among the five methods.

The paper has dealt with TS-FIS identification from small datasets. The numerical results obtained from sample data sets show a promising method proposed for TS-FIS model identification from small available historical data records.

# 6 Conclusion

We proposed in this paper a new method to identify effectively a standard T-S FIS designed for estimating bearings RUL, using historical data over an entire run-to-failure cycle collected from a small population of identical bearings. To offset the impact of the lack of data, we proposed a method based on the maximum likelihood estimation of mixture distribution analysis.

Within the proposed T-S FIS model identification approach, using the estimated priori probabilities of the fuzzy subsets and the estimated time clusters, the weighted least square estimation has allowed identifying the parameters of output membership functions of each fuzzy subset. Thus, the obtained parameters of the output membership function embed information associated with the population due to the priori probabilities and with the lifetime of bearings.

The numerical results show that the proposed approach yields a promising identification method of the T-S FIS model for bearing RUL estimation using small size datasets.

It makes sense that the proposed approach allows extracting and using complementary useful information from a small size dataset for bearings degradation model identification. The identification results obtained in section 5 are better than the other four methods based on the proper training dataset. A proper sampling method of the training datasets could ensure some overperformance of the results of the proposed method. Our future works will investigate the sampling process of training datasets of small size, together with an appropriate method of the FIS model parameters selection, identification and tuning, through an adaptive process. The original approach of feature selection also will be investigated for the selection of suitable features from the different existing time and frequency domain features for different types of bearings.